\documentclass[aps,reprint,groupedaddress,showpacs]{revtex4-1}

\usepackage{amsmath}
\usepackage{graphicx}
\usepackage{array}

\graphicspath{ {image/} }
\newcolumntype{L}[1]{>{\raggedright\arraybackslash}p{#1}}
\newcolumntype{C}[1]{>{\centering\arraybackslash}p{#1}}
\newcolumntype{R}[1]{>{\raggedleft\arraybackslash}p{#1}}

\begin{document}

\title{Transitions in synchronization states of model cilia through basal-connection coupling}
\author{Yujie Liu$^{1}$, Rory Claydon$^{2}$, Marco Polin$^{2,3*}$, and Douglas R. Brumley$^{1}$}
\email{d.brumley@unimelb.edu.au, M.Polin@warwick.ac.uk}

\affiliation{$^1$School of Mathematics and Statistics, The University of Melbourne, Victoria 3010, Australia. \\
$^2$Physics Department, University of Warwick, Gibbet Hill Road, Coventry CV4 7AL, UK. \\
$^3$Centre for Mechanochemical Cell Biology, University of Warwick, Gibbet Hill Road, Coventry CV4 7AL, UK.}

\begin{abstract}
Despite evidence for a hydrodynamic origin of flagellar synchronization between different eukaryotic cells, recent experiments have shown that in single multi-flagellated organisms, coordination hinges instead on direct basal body connections. The mechanism by which these connections leads to coordination, however, is currently not understood. Here we focus on the model biflagellate {\it Chlamydomonas reinhardtii}, and propose a minimal model for the synchronization of its two flagella as a result of both hydrodynamic and direct mechanical coupling. 
A spectrum of different types of coordination can be selected, depending on small changes in the stiffness of intracellular couplings. These include prolonged in-phase and anti-phase synchronization, as well as a range of multistable states induced by spontaneous symmetry breaking of the system. Linking synchrony to intracellular stiffness could lead to the use of flagellar dynamics as a probe for the mechanical state of the cell.
\end{abstract}

\maketitle

\section*{Introduction}

Cilia and flagella are structurally identical, whip-like cellular organelles employed by most eukaryotes for tasks ranging from sensing and locomotion of single cells \cite{Gray1955}, to directing embryonic development \cite{Nonaka2002} and driving cerebrospinal fluid flow \cite{Hagenlocher2013} in animals. Originally observed in 1677 by the Dutch pioneer Antoni van Leeuwenhoek, groups of motile cilia and flagella have a seemingly spontaneous tendency to coordinate their beating motion and generate large-scale patterns known as metachronal waves \cite{Knight_Jones:1954}. Coordination has often been proposed to provide an evolutionary advantage by improving transport and feeding efficiency \cite{Gueron:1999, Tam:2011, Taylor:1951, Mettot:2011, Michelin:2010fk, Ding2014}, although estimates of the magnitude of this effect are notoriously difficult.
Despite the uncertainty on its biological role, however, the universality of flagellar coordination is an empirical fact, and it suggests  the existence of a correspondingly general mechanism for synchronization. Mechanical forces, transmitted either by the surrounding fluid or internally through the cells, have often been proposed as responsible for this coordination \cite{Goldstein:2009vn,Uchida:2011kx,Geyer2013,Brumley2014,Klindt2017}.
Understanding how synchronization emerges could therefore highlight novel and potentially subtle roles played by physical forces in cell biology.  Here we develop a minimal model that links small changes in the mechanical properties of cells with the dynamics of their protruding flagella. In turn, this approach could  lead to coordination being used  as a probe to measure the internal mechanical state of a cell.

Reports of coordinated motion in nearby swimming sperm \cite{Rothschild:1949, Riedel2005} hint at the importance of hydrodynamic coupling. Hydrodynamic-led coordination of self-sustained oscillators, mimicking the active motion of cilia and flagella \cite{Bruot2016}, has been extensively investigated theoretically \cite{Vilfan:2006uq, Uchida:2011kx, Uchida:2012fk}, numerically \cite{Lagomarsino:2003zr, Lenz:2006fk, Wollin:2011}, and experimentally with colloidal rotors \cite{Brumley2016, Kotar:2010} and rowers \cite{DiLeonardo12, Uchida:2010ly}.
Despite the peculiar constraints of low-Reynolds number hydrodynamics, these studies suggest that hydrodynamic interactions can lead to ciliary coordination when coupled to either a phase-dependent driving force \cite{Uchida:2011kx,Uchida:2012fk}, or axonemal elasticity \cite{Kotar2013}. Indeed, hydrodynamic-mediated synchronization has been confirmed experimentally between pairs of eukaryotic flagella from different somatic cells from the green alga {\it Volvox carteri} \cite{Brumley2014}.  

Inter-cellular coordination of flagella, however, does not necessarily imply intra-cellular coordination, and therefore it is not {\it a priori} clear  whether hydrodynamic coupling is also responsible for the synchronization observed in individual multi-flagellated cells. 
Experimental studies of flagellar coordination within a single cell have focussed mainly on the biflagellate green alga {\it Chlamydomonas reinhardtii} \cite{Jeanneret2016, Goldstein2015} (CR), whose flagella are usually locked in a characteristic in-phase breaststroke motion. Early studies of flagellar coordination in CR \cite{Ringo1967, Ruffer1998, Ruffer1987} were recently refined \cite{Goldstein:2009vn, Polin:2009kx} and extended \cite{Goldstein:2011_emergence} using microfluidic devices and high-speed imaging of flagella \cite{Son2015}. These pointed at a fundamentally hydrodynamic origin for the observed synchronization, either through direct coupling or {\it via} a mechanism based on cell-body rocking \cite{Geyer2013}. However, a series of elegant novel experiments in CR and other flagellates challenged this view convincingly, showing instead that coordination requires the intracellular striated fibres that join flagellar basal bodies \cite{Quaranta2015, Wan2016}.  
Even though the precise mechanism by which direct connections affect flagellar coordination remains to be clarified \cite{Klindt2017}, the spontaneous transitions between extended in-phase (IP) and anti-phase (AP) beating in the CR mutant {\it ptx1} \cite{Leptos:2013}, already suggest that multiple synchronization states should be achievable through changes in the fibres' mechanical properties within the physiological range.

Here we propose a minimal model for flagellar dynamics for CR which can sustain both stable IP and stable AP states even in the absence of strong hydrodynamic coupling. Within this framework, the phase dynamics are determined principally by the mechanical state of the basal body fibres \cite{Wright1983}, with both types of coordination possible within a physiological range of fibres' stiffnesses \cite{MIT-course-notes}. 
The inclusion of hydrodynamic coupling leads to the emergence of a region in parameter space where non-trivial states can emerge as a result of spontaneous symmetry breaking through pitchfork bifurcations of limit cycles.

\begin{figure}[t]
\includegraphics[width=\columnwidth]{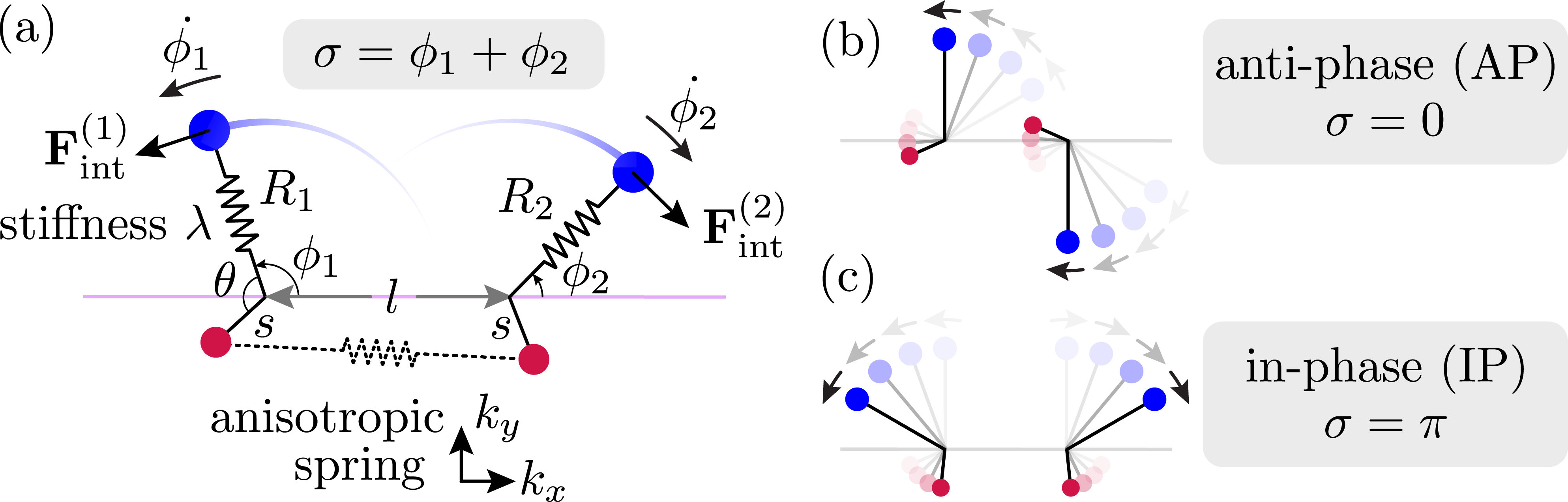}
\caption{Physical configuration. (a) Two external beads (blue), moving in the fluid, mimic the beating motion of flagellar filaments. Internal beads (red) represent the flagellar basal bodies, and are coupled through an anisotropic spring. (b) Anti-phase (AP) and (c) in-phase (IP) states.  \label{FIG1}}
\end{figure}

\section{Minimal model and leading order dynamics}
\label{sec:leading_model}

Figure~\ref{FIG1}a summarises our minimal model of flagella coupled through hydrodynamic and basal-body interactions. Following previous theoretical work \cite{Niedermayer:2008fk, Vilfan:2006uq}, and experimental measurements of flagellar flow fields and waveform elasticity \cite{Brumley2014}, the two flagella of {\it Chlamydomonas} are represented here by two spheres of radius $a$, immersed in a three dimensional fluid of viscosity $\mu$, and driven around circular orbits of radii $R_i$ ($i=1,2$) by constant tangential driving forces $\mathbf{F}_{\textrm{int}}^{(i)}$. 
Springs of stiffness $\lambda$ resist radial excursions from the equilibrium length $R_0$; and the magnitude of the internal driving forces, $F_{\textrm{int}}^{(i)} = 6\pi\mu aR_0\,\omega_i$  guarantees that, when isolated, the $i$-th oscillator will rotate at the intrinsic angular speed $\omega_i$. The orbits are centred along the $x$-axis, a distance $l$ apart, and lay along the $xy$ plane. Polar and radial coordinates $(\phi_i,R_i)$ define the oscillators' instantaneous positions around the centres of their respective orbits.

Both hydrodynamic and direct elastic interactions couple these minimal cilia. Hydrodynamic coupling is mediated by the fluid disturbance generated by each sphere's motion, modelled here as the flow from a point force. These interactions affect the instantaneous angular speeds $\dot{\phi}_i$ both directly, through a hydrodynamic torque, and indirectly by modifying the orbits' radii. For counter-rotating oscillators like those describing {\it Chlamydomonas} flagella, the resulting effective coupling will promote AP synchronization \cite{Niedermayer:2008fk, Ruffer1997, Leptos:2013, Klindt2017} (Fig.~\ref{FIG1}b).
In addition to external flagellar interactions, considerable evidence \cite{Quaranta2015, Wan2016} suggests that flagellar dynamics are strongly influenced by direct intracellular mechanical coupling, through striated fibres joining the basal bodies \cite{Ringo1967, Inouye1993} that can lead the system to IP synchrony (Fig.~\ref{FIG1}c). Intracellular connections are modelled here by introducing, for each oscillator, an auxiliary arm of fixed length $s \ (\ll R_0)$ at an angle $\theta$ ahead of the rotating sphere. The endpoints of these arms (red spheres in Fig.~\ref{FIG1}a) are coupled via an {\it anisotropic} elastic medium acting as elastic springs of stiffnesses $(k_x,k_y)$ and equilibrium lengths $(l,0)$ in the $x$ and $y$ directions respectively. This is intended to represent the intrinsically anisotropic structure of the fibre bundles connecting the basal bodies \cite{Wright1983}.
The equations of motion, derived in Supplementary Information \ref{SI_derivation}, follow from the requirement of zero net force and torque on each oscillator, and in the limit $R_0/l \ll 1$ can be approximated as (see Supplementary Information~\ref{SI_derivation})
\begin{align}
\begin{split}
& \dot{\phi}_1= \frac{R_0}{R_1}\omega_1+ \rho\frac{R_2}{R_1}\dot{\phi}_2J(\phi_1,\phi_2)+\rho\frac{\dot{R}_2}{R_1}K(\phi_1,\phi_2) \\
&+\frac{s^2}{\zeta R_1^2}(k_x+k_y)\left[G(\phi_1+\theta,\phi_2-\theta)-G(\phi_1+\theta,\phi_1+\theta)\right]
\\
& \dot{\phi}_2= \frac{R_0}{R_2}\omega_2+\rho\frac{R_1}{R_2}\dot{\phi}_1J(\phi_2,\phi_1)+\rho\frac{\dot{R}_1}{R_2}K(\phi_2,\phi_1) \\
&+\frac{s^2}{\zeta R_2^2} (k_x+k_y) \left[G(\phi_2-\theta,\phi_1+\theta)-G(\phi_2-\theta,\phi_2-\theta)\right]
\\
& \dot{R}_1= -\frac{\lambda}{\zeta}(R_1-R_0)+\rho R_2\dot{\phi}_2K(\phi_2,\phi_1)+\rho \dot{R}_2H(\phi_1,\phi_2)
\\
&\dot{R}_2 = -\frac{\lambda}{\zeta}(R_2-R_0)+\rho R_1\dot{\phi}_1K(\phi_1,\phi_2)+\rho \dot{R}_1H(\phi_2,\phi_1)
\end{split}
\label{eq:smallhydro}
\end{align}
where $\rho = 3a/8l$ ($\rho\ll1$ as $a\lesssim R_0$); $\zeta = 6\pi\mu a$ is the viscous drag coefficient of the rotating sphere; and
\begin{align}
\begin{split}
J(a,b)&=3\cos(a-b)-\cos(a+b),
\\
K(a,b)&=-3\sin(a-b)-\sin(a+b),
\\
H(a,b)&=3\cos(a-b)+\cos(a+b),
\\
G(a,b)&= \frac{1}{2} \frac{k_y-k_x}{k_x+k_y}\sin(a+b)-\frac{1}{2}\sin(a-b).
\end{split}
\label{eq:smallhydro_def}
\end{align}
In order to model the configuration typical of {\it Chlamydomonas} we will focus here on identical but counter-rotating oscillators, $\omega_1 = -\omega_2=\omega$. Parameter values are given in Table~\ref{table:params} unless otherwise specified.

\begin{table}
\begin{tabular}{| C{4.2cm} | C{1.5cm} | C{2.4cm} |}
\hline
{\it Variable} & {\it Symbol} & {\it Value} \\
\hline
model cilium radius \cite{Brumley2014} & $a$ & $0.75\ \mu$m \\
interflagellar spacing & $l$ & $15\ \mu$m \\
int./ext. bead angle & $\theta$ & 0 \\
external bead radius & $R_0$ & $5 \ \mu$m \\
internal bead radius & $s$ & $0.1 \ \mu$m \\
external spring stiffness \cite{Niedermayer:2008fk} & $\lambda$ & $4 \times 10^{-7} \ \text{Nm}^{-1}$ \\
viscosity of water & $\mu$ & $10^{-3} \ \text{Pa}\ \text{s}$ \\
ciliary beat frequency \cite{Goldstein:2009vn} & $f = \omega / 2 \pi$ & 50 Hz \\
\hline
\end{tabular}
\caption{Minimal model parameters used throughout, unless stated otherwise.}
\label{table:params}
\end{table}

\section{Fibres-only coupling}

\begin{figure}
\includegraphics[width=\columnwidth]{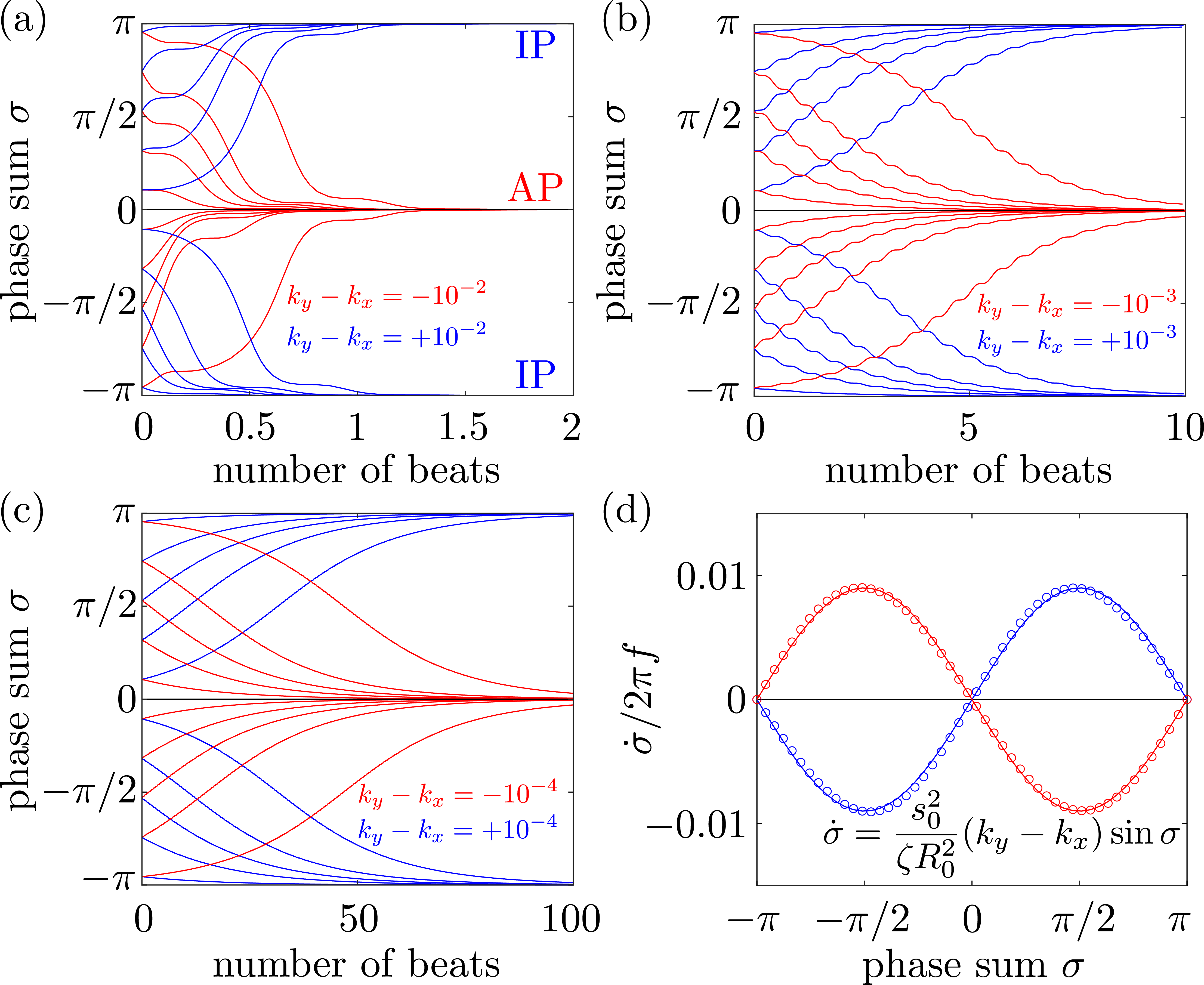}
\caption{Synchronization dynamics for pairs of model cilia in the absence of hydrodynamic interactions. Phase sum, $\sigma(t) = \phi_1(t) + \phi_2(t)$, for (a) $k_y-k_x= \pm 10^{-2}\ \text{Nm}^{-1}$, (b) $k_y-k_x= \pm 10^{-3}\ \text{Nm}^{-1}$ and (c) $k_y-k_x= \pm 10^{-4}\ \text{Nm}^{-1}$. In each case (starting with $\phi_1(0)=0$), IP (blue) and AP (red) synchronized states are obtained for $k_y > k_x$ and $k_y < k_x$, respectively, over a time-scale inversely proportional to $\text{max}(k_y,k_x)$. (d) Values of $\dot{\sigma}(\sigma)$ measured from numerical simulations (circles) compare favourably with Eq.~\eqref{eq:avgsigmanohydro} for sufficiently soft internal springs. Other model parameters as given in Table~\ref{table:params}.}
\label{FIG2}
\end{figure}

Despite the apparent simplicity, this minimal system displays rich dynamics, and it is therefore convenient to analyse its behaviour following steps of increasing complexity. Let us begin by considering the case in which hydrodynamic coupling is completely neglected. In this case $R_1=R_2=R_0$, and recasting the angular dynamics in terms of phase sum and difference $(\sigma,\delta) = (\phi_1+\phi_2,\phi_1-\phi_2)$ we obtain
\begin{align}
\dot{\sigma}&= (k_y-k_x)\frac{2s^2}{\zeta R_0^2}\,\sin^2\left(\frac{\delta}{2} + \theta\right)\,\sin(\sigma), \label{eq:fibresonly} \\
\dot{\delta}&=2\omega - \frac{s^2}{\zeta R_0^2}\left[(k_y+k_x) + (k_y-k_x)\cos(\sigma) \right]\sin(\delta +2\theta). \nonumber
\end{align}
Requiring that the maximal torque exerted on each oscillator by the internal springs is always smaller than that from the driving force ($\textrm{max}(k_x,k_y)s^2<\zeta\,R_0^2\omega$) guarantees that the cilia will always be beating ($\dot{\delta}>0$), and defines a physiologically plausible range for $k$'s, which in our case is $k_{x,y}\lesssim10^{-2}\,$Nm$^{-1}$. At the same time, unless $k_x=k_y$, the system will monotonically converge to either IP ($\sigma=\pi$) or AP ($\sigma=0$) synchrony, depending on whether $k_y$ is  larger or smaller than $k_x$. Figure~\ref{FIG2}a-c shows the convergence to either state, for a set of initial conditions and increasing internal stiffnesses. When $k_{x,y}\lesssim10^{-3}\,$Nm$^{-1}$, the phase sum evolves much faster than the difference, and $\sigma$ follows the approximate dynamics
\begin{equation}
\dot{\sigma} = (k_y-k_x)\frac{s^2}{\zeta R_0^2}\sin(\sigma),
\label{eq:avgsigmanohydro}
\end{equation}

as shown in Fig.~\ref{FIG2}d. The instantaneous and average phase speed profile can be solved analytically (see Supplementary Information~\ref{SI_phase_speed}). In the case of $k_x \text{ (in IP)}>k_y\text{ (in AP)}$, the model predicts a lower average phase speed in IP than in AP which is in qualitative agreement with experimental observation \cite{Leptos:2013}.
Recent studies argue that hydrodynamics plays a negligible role in flagellar synchronization for single cells \cite{Quaranta2015, Wan2016}. Without hydrodynamics, our model predicts that the effective interflagellar coupling should be given by $2\pi\epsilon=ks^2/\omega\zeta R_0^2$. Using the known value for {\it Chlamydomonas}, $|\epsilon|\simeq0.015$ \cite{Goldstein:2009vn,Leptos:2013}, we obtain $k\simeq 10^{-3}\,$Nm$^{-1}$ which translates to a Young's modulus $E\simeq 10^{5}\,$Pa for the bundle of striated fibres, when taking its length and thickness as $250\,$nm and $50\,$nm respectively \cite{Wright1983,Landau1986}.  This is a biologically plausible estimate, midway between the elastic modulus of relaxed skeletal muscle ($E\sim10^4\,$Pa) and elastin ($E\sim10^6\,$Pa) \cite{MIT-course-notes}.

\section{Stiff flagella hydrodynamics} \label{Section_stiff}

We begin now to include the effect of hydrodynamic interactions under the assumption of artificially stiff flagella, implemented in this section by increasing the radial spring stiffness to $\lambda=4\times10^{-6}\,\text{Nm}^{-1}$ ($10\times$ the value in Table~\ref{table:params}). Increasing $\lambda$ reduces the typical response time of the radial coordinate ($\zeta/\lambda$) and allows us to simplify the dynamics by assuming an instantaneous radial response \cite{Niedermayer:2008fk}. Then, to first order in the small parameter $\rho$, Eqs.~\eqref{eq:smallhydro} imply that $\sigma$ will obey
\begin{equation} 
\dot{\sigma} = \left[ (k_y-k_x)\frac{s^2}{\zeta R_0^2}-\frac{2\rho\omega^2\zeta}{\lambda}\right]\sin(\sigma),
\label{eq:sigma_smalhydrostiff}
\end{equation}
once the dynamics have been averaged over the fast variable $\delta$.
Within this approximation, hydrodynamics appears to simply shift the location of the transition between AP and IP states, determined by the steady state time average $\langle\sigma\rangle$, from $k_y=k_x$ to $k_y = k_x+2\rho(R_0/s)^2(\omega\zeta/\lambda)^2\lambda$. This is indeed confirmed by simulations of the full system for large $l$ (see Fig.~\ref{FIG4}b).

\begin{figure}[h!]
\includegraphics[width=\columnwidth]{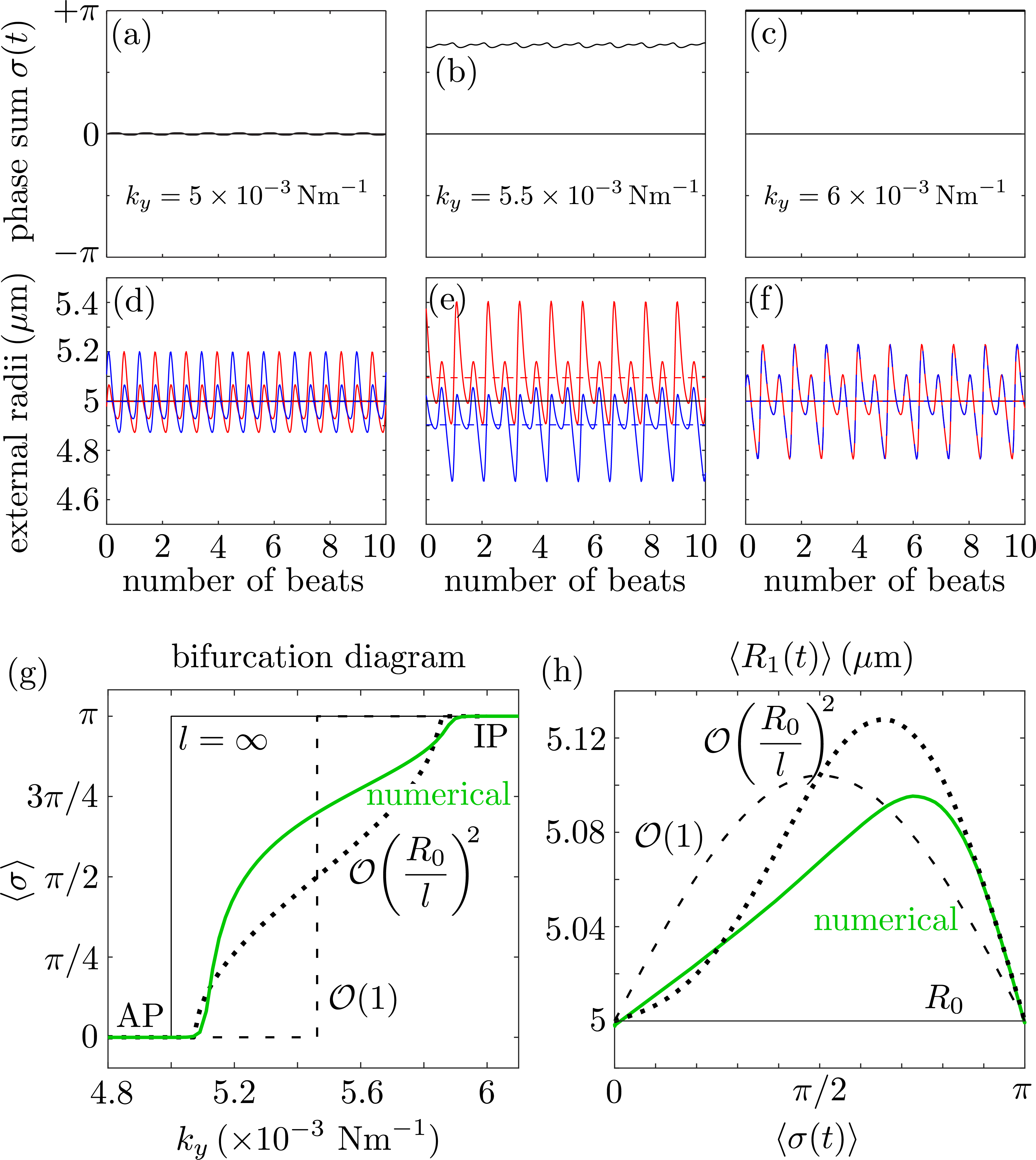}
\caption{Steady state solutions when radial dynamics are fast compared to the phase dynamics. This is achieved by setting $\lambda = 10 \lambda_0 = 4 \times 10^{-6} \ \text{Nm}^{-1}$. In each simulation $k_x = 5 \times 10^{-3} \ \text{Nm}^{-1}$. Other parameters shown in Table~\ref{table:params}. (a-c) Phase sum $\sigma(t) = \phi_1(t) + \phi_2(t)$ and (d-f) external rotor radii $R_i(t)$ are shown as functions of time for various values of $k_y$. (g) The mean value of phase sum, $\langle \sigma(t) \rangle$ and (h) external rotor radius $R_1(t)$ are computed from numerical solutions (green) and compared with analytical predictions of Eq.~\eqref{eq:sigma_smalhydrostiff} (dashed) and Eqs.~(\ref{eq:sigma_bifurcation},\ref{eq:R_bifurcation}) (dotted).}
\label{FIG3}
\end{figure}

A closer inspection, however, reveals more intriguing dynamics which become particularly evident for small inter-flagellar separations. Figure~\ref{FIG3}a-c shows the steady state dynamics of $\sigma$ for $l=15\,\mu$m, as $k_y$ is swept across the transition. Between AP and IP coordination (Figs.~\ref{FIG3}a,c) there is a distinct region of $k_y$-values for which the system synchronizes in a non-trivial intermediate state (Fig.~\ref{FIG3}b). This is accompanied by a permanent difference in the average values of the oscillators' radii (Fig.~\ref{FIG3}e), with the asymmetry depending on which of the equally probable signs of $\langle\sigma\rangle$ is chosen by the system.
Figure~\ref{FIG3}g shows the full positive branch of $\langle\sigma\rangle$ as $k_y$ is swept between AP and IP values (simulations: green solid line). This can be compared to the leading-order behaviour with and without hydrodynamics (black dashed and solid lines); and the one predicted by refining Eq.~\eqref{eq:sigma_smalhydrostiff} to next-to-leading order in $R_0/l$  (black dotted line; see Supplementary Information~\ref{SI_HOT})
\begin{equation}
\dot{\sigma} = \left[ (k_y-k_x)\frac{s^2}{\zeta R_0^2}-\frac{2\rho\omega^2\zeta}{\lambda}\left(1-\frac{15}{2}\frac{R_0^2}{l^2}\cos(\sigma)\right)\right]\sin(\sigma).
\label{eq:sigma_bifurcation}
\end{equation}
The semi-quantitative agreement between the simulated and predicted dependence of the steady-state $\langle\sigma\rangle$ extends also to the $k_y$-dependence of the time-averaged radii (Fig.~\ref{FIG3}h), which follows in the same approximation
\begin{equation}
R_{1,2} = R_0\left[1\pm\frac{\rho\omega\zeta}{\lambda}\left(1-\frac{15}{2}\frac{R_0^2}{l^2}\cos(\sigma)\right)\sin(\sigma)\right].
\label{eq:R_bifurcation}
\end{equation}
\begin{figure}[t]
\includegraphics[width=\columnwidth]{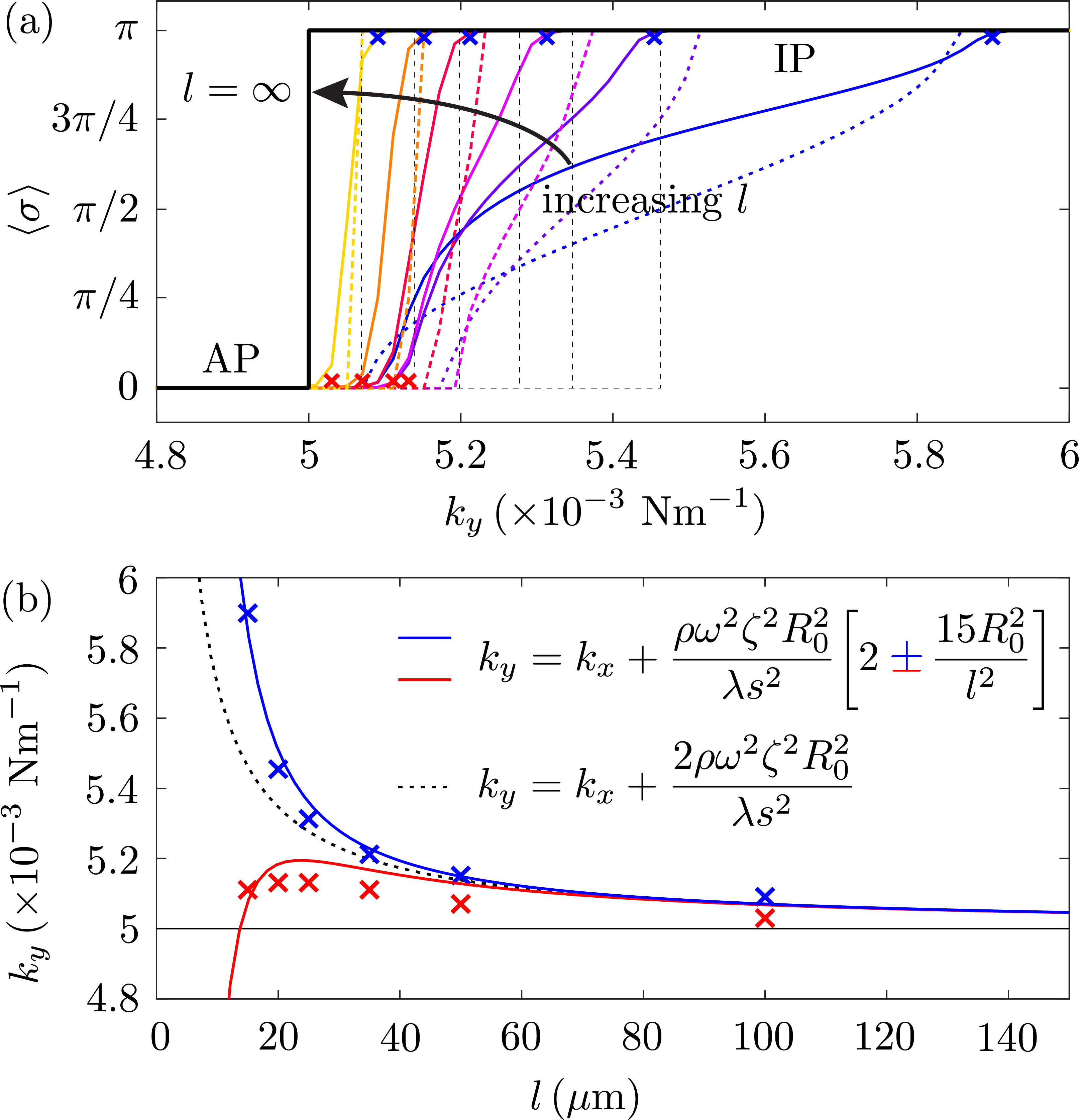}
\caption{Transitions between synchronization states mediated by basal body coupling. Radial dynamics are again fast compared to phase dynamics ($\lambda = 10 \times \lambda_0 = 4 \times 10^{-6} \ \text{Nm}^{-1}$). (a) Steady state $\langle \sigma(t) \rangle$ as a function of $k_y$ for various ciliary spacings ($l=15$, 20, 25, 35, 50, 100$\, \mu$m). Numerical solutions (smooth, coloured) are shown alongside analytical predictions to leading order (dashed, black) and second order in $R_0/l$ (dashed, coloured). The transition zone boundaries are quantified by $\langle \sigma(t) \rangle = 0.02 \pi$ (red crosses) and $\langle \sigma(t) \rangle = 0.98 \pi$ (blue crosses) respectively. (b) These are shown as a function of $l$, together with leading order (black) and second order in $R_0/l$ (coloured) analytical predictions.}
\label{FIG4}
\end{figure}
Figure~\ref{FIG4}a shows that the agreement extends across the full range of separations $l\geq 15\,\mu$m, with particularly accurate estimates for the values of $k_y$ marking the beginning and end of the transition (Fig.~\ref{FIG4}b). These results suggest that the simple expression in Eq.~\eqref{eq:sigma_bifurcation} captures the essential features of the dynamics, and can therefore be used to analyse the nature of the transition. For small deviations $\delta k$ in $k_y$ around the transition from AP, Eq.~\eqref{eq:sigma_bifurcation} can be approximated as 
\begin{equation}
\dot{\sigma}\simeq \left(\delta k\frac{s^2}{\zeta R_0^2}\right)\sigma-\left(\frac{15}{2}\frac{\rho\omega^2\zeta}{\lambda}\frac{R_0^2}{l^2}\right)\sigma^3,
\label{eq:sigma_pitchfork}
\end{equation}
which therefore suggests that the emergence of non-trivial coordination follows a supercritical pitchfork bifurcation \cite{Strogatz:Nonlinear_Dynamics}. A similar argument leads to an equivalent conclusion for the bifurcation from IP as $k_y$ is decreased. AP and IP domains are therefore bounded by a pitchfork bifurcation of limit cycles.

Within the intermediate regime, the system becomes naturally bistable through a spontaneous symmetry breaking from a state where both oscillators follow the same limit cycle to one where they sustain a stable difference in their average oscillation amplitudes. Transitions between homogeneous and inhomogeneous oscillation states, and bistability, have only recently been discovered in pairs of coupled limit cycle oscillators, also as a consequence of non-equilibrium symmetry-breaking pitchfork bifurcations \cite{Rohm2017, Banerjee2018}. Here we discover their emergence in a simple model of hydrodynamic- and basal-body-coupled flagella.
Intermediate equilibrium states appear when the internal elastic interaction promoting IP coordination is approximately compensated by the leading order hydrodynamic coupling favouring AP, amplifying the importance of higher-order hydrodynamic effects. 
In this parameter range, the system becomes naturally bistable through a spontaneous symmetry breaking from a state where both oscillators follow the same limit cycle to one where they sustain a stable difference in their average oscillation amplitude.
Interestingly, the permanent difference in the average radii of the two oscillators after the bifurcation could be easily interpreted by an observer as a difference in intrinsic frequency. The rotors would then appear intrinsically different despite in fact being identical. 
Eventually, a sufficient increase of the internal stiffness can overcome the antagonistic effect of hydrodynamic interactions at any given separation $l$, and drive the system to a stable IP state. 

\section{The full model} \label{Section_full_model}

\begin{figure}[htp]
\begin{center}
\includegraphics[width=\columnwidth]{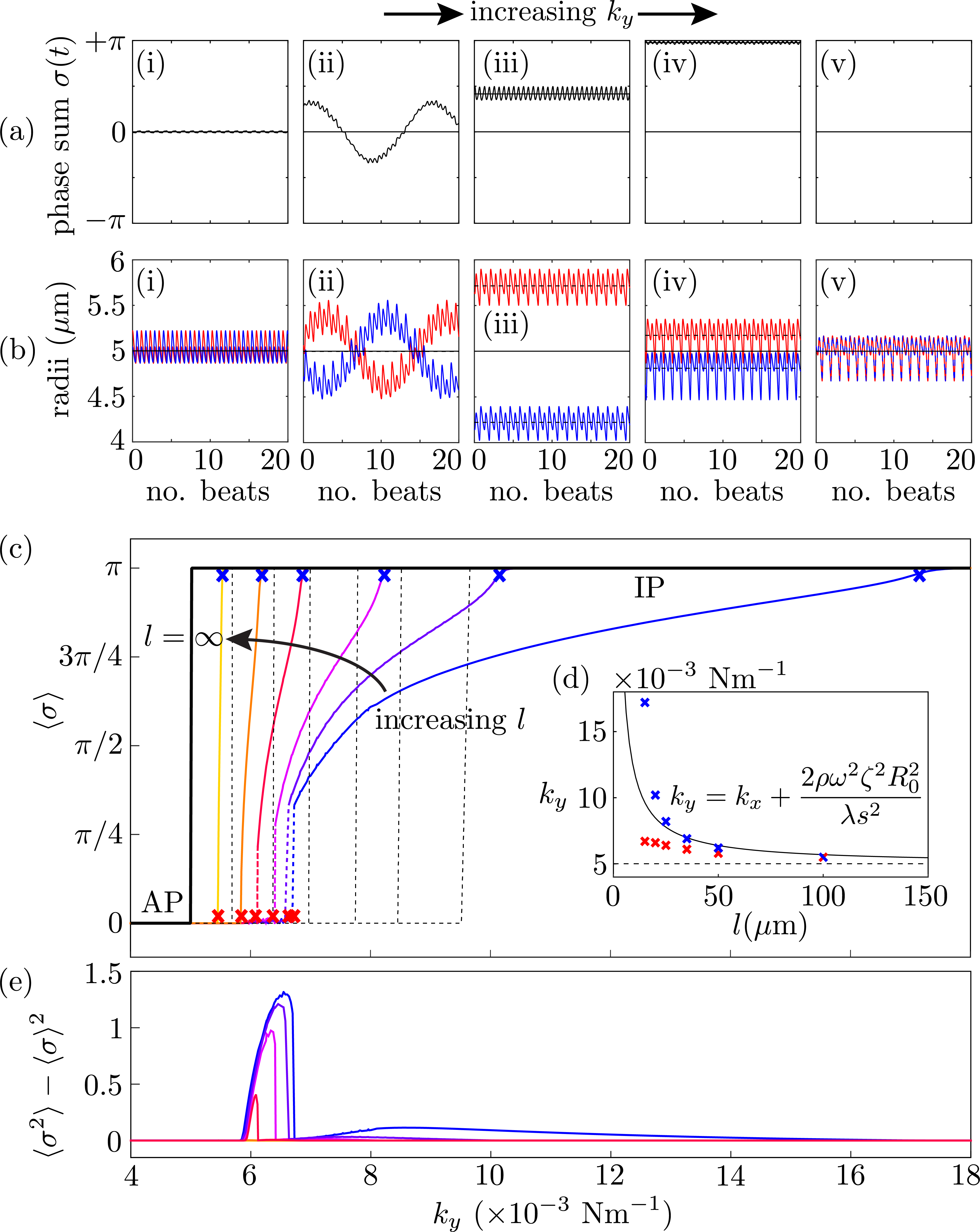}
\end{center}
\caption{Basal body coupling and hydrodynamic interactions. (a) Steady state phase sum $\sigma(t) = \phi_1(t) + \phi_2(t)$ and (b) external radii $R_1(t)$ and $R_2(t)$. Results are shown for $k_y = 0.0055$, 0.006, 0.007, 0.017, 0.020 $ \text{Nm}^{-1}$ with $k_x = 0.005 \ \text{Nm}^{-1}$ and $l = 15\, \mu$m. For intermediate values of $k_y$, the external rotors possess a permanent difference in their intrinsic frequencies. (c) Value of $\langle \sigma(t) \rangle$ as a function of $k_y$ for various values of $l$ (results shown for $l=15$, 20, 25, 35, 50, 100$\, \mu$m). The transition zone boundaries are quantified by $\langle \sigma(t) \rangle = 0.02 \pi$ (red crosses) and $\langle \sigma(t) \rangle = 0.98 \pi$ (blue crosses) respectively. (d) Measured boundaries compared with far-field analytical results. (e) The variance of $\sigma(t)$ reveals large excursions in the phase sum prior to the bifurcation (see also panel (a)ii).}
\label{FIG5}
\end{figure}

We conclude by looking at the full system with realistic parameters throughout (see Table~\ref{table:params}). 
In this case, radial and phase dynamics have comparable timescales ($\zeta\omega/2\pi\lambda\sim 2$) and the radii cannot be considered as approximately slaved to the phases anymore. Together with the sizeable radial deformations ($\delta R/R_0\sim R_0/2l$ here and $\sim0.2$ for {\it Chlamydomonas}-like $l=15\,\mu$m) this results in a complex interplay between radial and phase variables, and implies the need to consider the full system of governing equations (see Supplementary Information~\ref{SI_derivation}). These will be explored here through numerical simulations only.

Figure~\ref{FIG5}a,b shows a representative set of curves for a $k_y$ sweep with $k_x=5\times10^{-3}\,\text{Nm}^{-1}$ and $l=15\,\mu$m. Similarly to the case of stiffer flagella, low and high values of $k_y$ correspond respectively to AP (panels (i)) and IP (panels (v)) synchronization, and in these states the oscillators follow the same dynamics. 
The average phase speed, however, is observed to be lower in IP than in AP (a difference of $\sim 13\%$ between $k_y=0$ Nm$^{-1}$ and $k_y=0.02$ Nm$^{-1}$), in qualitative agreement with experimental observations of CR mutants \cite{Leptos:2013} (see Supplementary Information~\ref{SI_phase_speed}). 
In between, there is a range of $k_y$ values for which the system synchronizes around intermediate values of $\langle\sigma\rangle$, with the two oscillators following again different limit cycles (panels (iii,iv)).
A new state, however, appears as $k_y$ approaches the symmetry breaking transition from the AP side (panels (ii)). Despite corresponding formally to AP ($\langle\sigma\rangle=0$), the system displays symmetric excursions in the relative phase difference which are long lived and not much smaller than $\pi$. In this state, the oscillators spend most of their time at values of $\sigma$ far from 0, and the null average is only guaranteed by the symmetry of the dynamics. Although the system oscillates here by about $\pi/3$, amplitudes  $\sim\pi$ can be easily obtained just by increasing $a$ (Supplementary Information Fig.~\ref{FIG_S2}). In this condition the system will not appear synchronized in AP at all, but will rather be continuously alternating between IP at $\pi$ and IP at $-\pi$.
Figure~\ref{FIG5}c,e shows that this situation is typical for all the separations  displaying a discontinuous, rather than continuous, transition out of the AP state (here all $l\leq35\,\mu$m). In the IP case, instead, the transition maintains its continuous nature throughout, and in fact the bifurcation point is still well predicted by the first order expression from Eq.~\eqref{eq:sigma_smalhydrostiff} (see inset). 

\begin{figure}[h!]
\begin{center}
\includegraphics[width=\columnwidth]{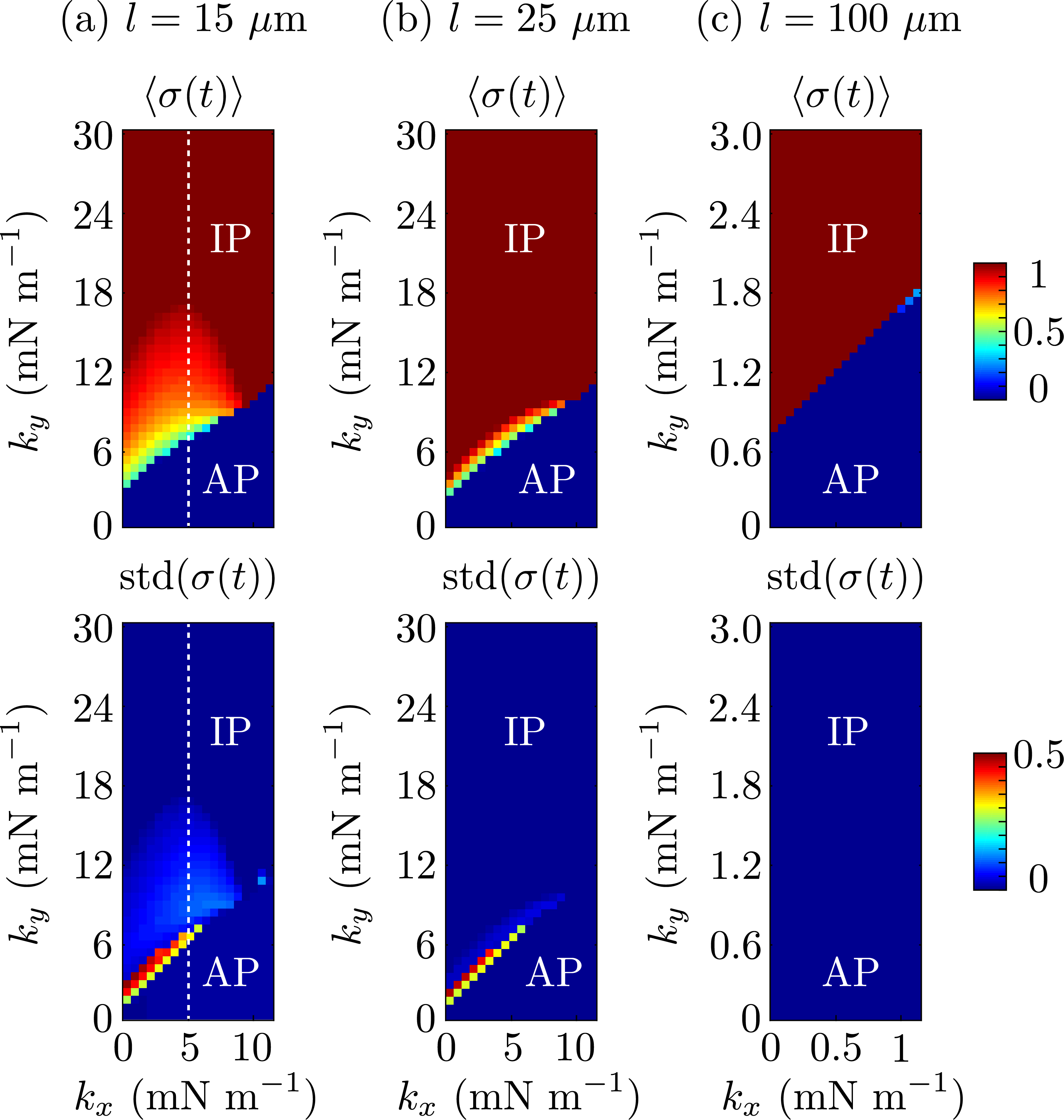}
\end{center}
\caption{Transition between AP and IP states is mediated by internal spring constants. Mean and standard deviation of $\sigma(t)$ as functions of $k_x$ and $k_y$ for external rotor separations (a) $l=15 \ \mu$m, (b) $l=25 \ \mu$m, and (c) $l=100 \ \mu$m. White dotted lines correspond to the blue bifurcation plot in Fig.~\ref{FIG5}(c). For small $k_x$, the system is capable of supporting intermediate phase-locked states, with $0 < \langle \sigma(t) \rangle < \pi$. However, for larger $k_x$, an abrupt transition between AP and IP occurs.}
\label{FIG6}
\end{figure}

From the AP side, discontinuous transitions are always preceded by a region of $k_y$ values where the system displays a large excursion dynamics, which therefore acts as a predictor of the impending discontinuity \cite{Scheffer2009}. 
The presence of the discontinuity in $\langle\sigma\rangle$, and the preceding large fluctuations, depend on both the separation $l$ and  $k_x$, as shown in Fig.~\ref{FIG6} for a $(k_x,k_y)$ parameter sweep. 
For realistic inter-flagellar separation ($l=15\,\mu$m), Fig.~\ref{FIG6}a shows that the region of discontinuous transition out of AP, marked by the large standard deviation of $\sigma(t)$, exists only for $k_x\lesssim 9 \times 10^{-3}\,\text{Nm}^{-1}$. Above this value, the bifurcation changes its nature and becomes continuous but sharp.
At the slightly larger separation of $l=25 \mu$m, the width of the extended transition zone observed previously for $k_x\lesssim 9 \times 10^{-3}\,\text{Nm}^{-1}$ is reduced (see Figs.~\ref{FIG6}b, \ref{FIG5}c).
Further increasing the separation to $l=100\,\mu$m reduces hydrodynamic forces by an order of magnitude compared to the $l=15\,\mu$m case, and for all the $k_x$ values explored, the system follows a sharp continuous transition from AP to IP as $k_y$ is increased. 

Although exploring in detail the nature of these bifurcations is beyond the scope of the present work, clear similarities with the  case of stiff flagella suggest strongly that the qualitative nature of the continuous transition is the same in the two cases. We expect therefore the continuous transitions to be supercritical pitchfork bifurcations of limit cycles, inducing the observed symmetry breaking in the system (Fig.~\ref{FIG5}a (iii-iv)).
For $l\leq35\,\mu$m, the emergence of a discontinuity in $\langle\sigma\rangle$ implies that decreasing $k_x$ can change the nature of the transition. Looking closely at the discontinuous case, we find that there is an extended region of overlap between the $\langle\sigma\rangle=0$ and the intermediate $\langle\sigma\rangle$ branches (see Supplementary Information \ref{SI_coexistence}). This is typical of a catastrophe-like transition which, given the $\sigma\leftrightarrow-\sigma$ symmetry of the system, is likely to be a subcritical pitchfork bifurcation.

Coexistence of three different states, all of which are locally stable for the dynamics, means that the system displays multi-stability: presence of noise might then induce the system to jump between these locally stable states and therefore alternate between periods of AP synchronization an other non-trivial types of coordination, with transitions dictated by escape rate arguments \cite{Thompson2011, Miller2012, Herbert2017}.

\section{Conclusions}

While mechanisms for hydrodynamic-led synchronization of cilia and flagella have been extensively studied \cite{Elgeti:2013, Brumley2014, Vilfan:2006uq, Uchida:2011kx, Uchida:2012fk, Lagomarsino:2003zr, Lenz:2006fk, Wollin:2011, Brumley2016, Kotar:2010}, the impact of direct intra-cellular connections on flagellar dynamics is only starting to be recognised \cite{Quaranta2015, Wan2016, Klindt2017}. Here we have extended a simple and popular minimal model for the hydrodynamically interacting flagella pair of {\it Chlamydomonas} to account for intracellular mechanical coupling. The clearly anisotropic ultrastructure of striated fibres \cite{Wright1983} is mirrored in the use of a non-isotropic elastic interaction between the oscillators ($k_x\neq k_y$), and results in a phase-phase coupling that can promote {\it by itself} either IP or AP synchronization, within a biologically plausible range of Young's moduli. Transitions would then result simply through changes in the relative magnitude of $k_x$ and $k_y$. Given that intracellular calcium can control the contraction of striated fibres in {\it Chlamydomonas} \cite{Harris2009}, we hypothesise that the transitions in coordination observed experimentally could be the result of localised apical variations in cytoplasmic $[\text{Ca}^{2+}]$ \cite{Collingridge2013, Wheeler2008a, Wheeler2008b}.
Natural extensions of this model to amplitude-phase coupling do not influence the leading order coordination dynamics (see Supplementary Information~\ref{SI_derivation} for brief discussion) and have been omitted here. 
In-phase coordination has recently been proposed to result from a different nonlinear interplay between hydrodynamic and intracellular mechanical coupling \cite{Klindt2017}, with AP due to either one of them operating in isolation. However, several experimental observations, from the absence of phase-locking in mutants lacking striated fibres \cite{Quaranta2015, Wan2016}, to the complex synchronization observed in multi-flagellated algae \cite{Wan2016}, suggest that hydrodynamics plays in fact only a minimal role in this system. The model introduced here can sustain both AP and IP states without the need for external coupling, through a mechanism potentially under the direct control of the cell.  Unequal tightening of the different fibres joining the basal bodies of cells with more than two flagella could then lead to the complex synchronization patterns observed experimentally \cite{Wan2016}. 
Yet, subtle hydrodynamic effects can exist, and need to be investigated through dedicated experiments blocking fluid-mediated coupling between flagella. These are currently underway.

With coordination of cilia and flagella within single cells being sensitive to the direct intracellular coupling between the filaments, we believe that better understanding its emergence will eventually enable synchronization to be used as a new and sensitive probe for the intracellular mechanical state of a cell.

\bibliography{Liu_etal.bbl}

\newpage

\setcounter{figure}{0}
\renewcommand{\thefigure}{S\arabic{figure}}
\setcounter{equation}{0}
\renewcommand{\theequation}{S\arabic{equation}}

\setcounter{section}{0}
\renewcommand{\thesection}{S\arabic{section}}

\onecolumngrid

\vspace{22pt}

\begin{center}
\begin{LARGE}
{\sc Supplementary Information}
\end{LARGE}
\end{center}

\vspace{11pt}

\section{Derivation of governing equations} \label{SI_derivation}

In this section we derive the full equations of motion for arbitrary $R_0/l$. The beads on the left and right hand side of Fig.~\ref{FIG1}a are referred to as 1 and 2 respectively. We use the centre of the circular orbit of bead 1 as the origin of reference. The polar unit vectors for external and internal beads (1 and 2) are given by
\begin{align}
\begin{split}
\mathbf{e}_{r_{1,2}}&=(\cos\phi_{1,2},\sin\phi_{1,2}), \\
\mathbf{e}_{\phi_{1,2}}&=(-\sin\phi_{1,2},\cos\phi_{1,2}), \\
\mathbf{w}_{r_{1,2}}&=(\cos(\phi_{1,2}\pm\theta),\sin(\phi_{1,2}\pm\theta)), \\
\mathbf{w}_{\phi_{1,2}}&=(-\sin(\phi_{1,2}\pm\theta),\cos(\phi_{1,2}\pm\theta)).
\end{split} \label{fulleq:unitvec}
\end{align}
The equations for bead 1 will be derived explicitly, and the corresponding equations for bead 2 obtained by symmetry. The two internal beads (see Fig.~\ref{FIG1}a) are connected by an anisotropic spring of stiffness $(k_x,k_y)$ and equilibrium length $(l,0)$. The force exerted by internal bead 2 on internal bead 1 through this spring is
\begin{align}
\begin{split}
f_x&=k_xs(\cos(\phi_2-\theta)-\cos(\phi_1+\theta)),
\\
f_y&=k_ys(\sin(\phi_2-\theta)-\sin(\phi_1+\theta)).
\end{split} \label{fulleq:forcexy}
\end{align}
In polar coordinates, we are only interested in the tangential force as the there is no radial freedom for the internal beads in the model.
\begin{align}
\begin{split}
f_{\phi_1}&=(\mathbf{w}_{\phi_1}\cdot\mathbf{e}_x)f_x+(\mathbf{w}_{\phi_1}\cdot\mathbf{e}_y)f_y=-f_x\sin(\phi_1+\theta)+f_y\cos(\phi_1+\theta).
\end{split} \label{fulleq:forcepol}
\end{align}
Substituting Eq.~\eqref{fulleq:forcexy} into Eq.~\eqref{fulleq:forcepol} yields
\begin{align}
\begin{split}
f_{\phi_1}&= (k_x+k_y) s \big[ G(\phi_1+\theta,\phi_2-\theta)- G(\phi_1+\theta,\phi_1+\theta)] ,
\end{split} \label{fulleq:fp}
\end{align}
where for angles $a$ and $b$,
\begin{align}
\begin{split}
G(a,b)&=\frac{1}{2} \frac{k_y-k_x}{k_x+k_y}\sin(a+b)- \frac{1}{2} \sin(a-b).
\end{split} \label{fulleq:def}
\end{align}
To obtain the governing equations of the system, the following assumptions are made:
\begin{enumerate}
\item The torque of the system is balanced.
\item The radial force of the system is balanced for each bead.
\end{enumerate}
This essentially assumes that the system can respond to the external force instantaneously and reach the equilibrium. The motion of the external beads creates hydrodynamic disturbances which are modelled as Stokeslets in an unbounded fluid \citep{Niedermayer:2008fk}. To express the Stokeslet flow field produced by external bead 2, but located at external bead 1, we define the effective Stokeslet strength $\mathbf{f}$, which is proportional to the velocity of external bead 2. The relative positions of the external beads is given by $\mathbf{r}_{12}=\mathbf{r}_1-\mathbf{r}_2$.
\begin{align}
\begin{split}
\mathbf{f}&=\frac{3a}{4} R_2\dot{\phi}_2\mathbf{e}_{\phi_2}+\frac{3a}{4}\dot{R}_2\mathbf{e}_{r_2}\\
&=(-l+R_1\cos\phi_1-R_2\cos\phi_2)\mathbf{e}_x+(R_1\sin\phi_1-R_2\sin\phi_2)\mathbf{e}_y
\\
\frac{4}{3a}\mathbf{f}\cdot \mathbf{r}_{12}
&=l(R_2\dot{\phi}_2\sin\phi_2-\dot{R}_2\cos\phi_2)+R_1\dot{R}_2\cos(\phi_1-\phi_2)+R_1R_2\dot{\phi}_2\sin(\phi_1-\phi_2)-R_2\dot{R}_2
\\
r_{12}^2
&=l^2-2l(R_1\cos\phi_1-R_2\cos\phi_2)+R_1^2+R_2^2-2R_1R_2\cos(\phi_1-\phi_2)
\end{split} \label{fulleq:def1}
\end{align}
\begin{itemize}
\item[(1)]\textbf{Torque Balance}\\
The torque balance for the overdamped bead 1 system is similar to \citep{Niedermayer:2008fk}, but with an additional term. The internal driving force from the flagellar motors is $\mathbf{F}^{(1,2)}_{\text{int}}=\zeta R_0\omega_{1,2}$.
\begin{align}
R_1\dot{\phi}_1-\mathbf{e}_{\phi_1}\cdot\left(\frac{\mathbf{f}}{r_{12}}+\frac{\mathbf{r}_{12}(\mathbf{f}\cdot\mathbf{r}_{12})}{r_{12}^3}\right)=R_0\omega_1+\frac{sf_{\phi_1}}{\zeta R_1}. \label{fulleq:torque}
\end{align}
The LHS of Eq.~\eqref{fulleq:torque} can be expanded using Eq.~\eqref{fulleq:def1} to obtain
\begin{align}
\begin{split}
R_1\dot{\phi}_1-\frac{3a}{4r_{12}}\left[\frac{M^{+}_{12}A^{+}_{12}}{r_{12}^2}+R_2\cos(\phi_1-\phi_2)\right]\dot{\phi}_2-\frac{3a}{4r_{12}}\left[\frac{M^{+}_{12}B^{+}_{12}}{r_{12}^2}+\sin(\phi_2-\phi_1)\right]\dot{R}_2.
\end{split} \label{fulleq:p1}
\end{align}
The equations for bead 2 can be obtained by swapping the labels (1 and 2) and changing the sign of $l$. Here we make use of the translational invariance property of the system. This is equivalent to translating the whole system horizontally by $l$ such that the centre of orbit of bead 2 coincides the origin of reference frame
\begin{align}
R_2\dot{\phi}_2-\frac{3a}{4r_{12}}\left[\frac{M^{-}_{21}A^{-}_{21}}{r_{12}^2}+R_1\cos(\phi_2-\phi_1)\right]\dot{\phi}_1-\frac{3a}{4r_{12}}\left[\frac{M^{-}_{21}B^{-}_{21}}{r_{12}^2}+\sin(\phi_1-\phi_2)\right]\dot{R}_1, \label{fulleq:p2}
\end{align}
where
\begin{align}
\begin{split}
M^{\pm}_{ij}&=\pm l\sin\phi_i+R_j\sin(\phi_i-\phi_j), \\
A^{\pm}_{ij}&=\pm lR_j\sin\phi_j+R_iR_j\sin(\phi_i-\phi_j), \\
B^{\pm}_{ij}&=\mp l\cos\phi_j+R_i\cos(\phi_i-\phi_j)-R_j.
\end{split} \label{fulleq:def2}
\end{align}
\item[(2)]\textbf{Radial Force Balance}\\
Similarly, with the external spring of stiffness $\lambda$, the radial force balance is given by the expression
\begin{align}
\dot{R}_1-\mathbf{e}_{r_1}\cdot\left(\frac{\mathbf{f}}{r_{12}}+\frac{\mathbf{r}_{12}(\mathbf{f}\cdot\mathbf{r}_{12})}{r_{12}^3}\right)=-\frac{\lambda}{\zeta}(R_1-R_0). \label{fulleq:radial}
\end{align}
Although the internal beads contribute indirectly, the above equation does not involve an explicit contribution from their motion. Using Eq.~\eqref{fulleq:def1} we can expand the LHS of Eq.~\eqref{fulleq:radial}
\begin{align}
\dot{R}_1-\frac{3a}{4r_{12}}\left[\frac{N^{+}_{12}A^{+}_{12}}{r_{12}^2}+R_2\sin(\phi_1-\phi_2)\right]\dot{\phi}_2-\frac{3a}{4r_{12}}\left[\frac{N^{+}_{12}B^{+}_{12}}{r_{12}^2}+\cos(\phi_1-\phi_2)\right]\dot{R}_2.  \label{fulleq:r1}
\end{align}
By the same symmetry argument, we obtain the radial equation for bead 2
\begin{align}
\dot{R}_2-\frac{3a}{4r_{12}}\left[\frac{N^{-}_{21}A^{-}_{21}}{r_{12}^2}+R_1\sin(\phi_2-\phi_1)\right]\dot{\phi}_1-\frac{3a}{4r_{12}}\left[\frac{N^{-}_{21}B^{-}_{21}}{r_{12}^2}+\cos(\phi_2-\phi_1)\right]\dot{R}_1, \label{fulleq:r2}
\end{align}
where
\begin{align}
N^{\pm}_{ij}&=\mp l\cos\phi_i+R_i-R_j\cos(\phi_i-\phi_j). \label{fulleq:def3}
\end{align}
\end{itemize}
From Eqs.~\eqref{fulleq:p1}-\eqref{fulleq:p2} and Eqs.~\eqref{fulleq:r1}-\eqref{fulleq:r2}, the governing equations are therefore
\begin{align}
\begin{split}
&R_1\dot{\phi}_1-\frac{3a}{4r_{12}}\left[\frac{M^{+}_{12}A^{+}_{12}}{r_{12}^2}+R_2\cos(\phi_1-\phi_2)\right]\dot{\phi}_2-\frac{3a}{4r_{12}}\left[\frac{M^{+}_{12}B^{+}_{12}}{r_{12}^2}+\sin(\phi_2-\phi_1)\right]\dot{R}_2=R_0\omega_1
+\frac{sf_{\phi_1}}{\zeta R_1},
\\
&R_2\dot{\phi}_2-\frac{3a}{4r_{12}}\left[\frac{M^{-}_{21}A^{-}_{21}}{r_{12}^2}+R_1\cos(\phi_2-\phi_1)\right]\dot{\phi}_1-\frac{3a}{4r_{12}}\left[\frac{M^{-}_{21}B^{-}_{21}}{r_{12}^2}+\sin(\phi_1-\phi_2)\right]\dot{R}_1=R_0\omega_2
+\frac{sf_{\phi_2}}{\zeta R_2},
\\
&\dot{R}_1-\frac{3a}{4r_{12}}\left[\frac{N^{+}_{12}A^{+}_{12}}{r_{12}^2}+R_2\sin(\phi_1-\phi_2)\right]\dot{\phi}_2-\frac{3a}{4r_{12}}\left[\frac{N^{+}_{12}B^{+}_{12}}{r_{12}^2}+\cos(\phi_1-\phi_2)\right]\dot{R}_2=-\frac{\lambda}{\zeta}(R_1-R_0),
\\
&\dot{R}_2-\frac{3a}{4r_{12}}\left[\frac{N^{-}_{21}A^{-}_{21}}{r_{12}^2}+R_1\sin(\phi_2-\phi_1)\right]\dot{\phi}_1-\frac{3a}{4r_{12}}\left[\frac{N^{-}_{21}B^{-}_{21}}{r_{12}^2}+\cos(\phi_2-\phi_1)\right]\dot{R}_1=-\frac{\lambda}{\zeta}(R_2-R_0),
\end{split} \label{fulleq}
\end{align}
where $f_{\phi_2}$ is $f_{\phi_1}$ in Eq.~\eqref{fulleq:fp} but with the exchange $\phi_1+\theta\leftrightarrow\phi_2-\theta$. \\

The set of governing equations is non-linear, and it is useful to simplify the system while preserving some of its important features. One way is to take the small hydrodynamic limit of Eq.~\eqref{fulleq}, i.e. as $l\gg R_0$, ignore $O(R_{1,2}/l)$. Under this assumption, we obtain Eq.~\eqref{eq:smallhydro} in the paper. If the timescale for changes in $\sigma = \phi_1 + \phi_2$ is long compared to the timescale for $\delta=\phi_1-\phi_2$ (as is the case in section~\ref{Section_stiff}), the leading order equation (Eq.~\eqref{eq:sigma_smalhydrostiff}) for the phase sum $\sigma$ can be obtained by summing the leading order contributions from each term in the equations for $\phi_1$ and $\phi_2$. This gives the hydrodynamic interaction term in \citep{Niedermayer:2008fk} plus the leading order term from the anisotropic spring interaction, and is shown in Eq.~\eqref{eq:avgsigmanohydro} of the main paper. 

This model has the advantage that it can be easily generalized so that the internal basal coupling involves both the phase and amplitude of the external rotors. One possible way is to enable radial freedom to the internal beads, which can be coupled with the amplitude of the external oscillators via a larger spring that connects the two beads. However, it can be shown that under the assumptions that (i) the internal spring constant is stiff compared to the sum of external and large spring constants, and (ii) the internal bead radius is small compared to the external bead radius, this internal radial freedom does not contribute to the leading order dynamics in Eq.~\eqref{eq:sigma_smalhydrostiff}.

\section{Derivation of Next Order Geometric Higher Order Term} \label{SI_HOT}

In this section, we present the derivation of next-to-leading order term in $R_0/l$ in Eqs.~\eqref{eq:sigma_bifurcation}-\eqref{eq:R_bifurcation}. We aim at geometric higher order terms ($O(R/l)$) in the radial equation for $R_1$ in  Eq.~\eqref{fulleq}. The case for $R_2$ can be obtained by symmetry. Since we are interested in the leading order DC component of $R_1$ at equilibrium, we can drop the term proportional to $\dot{R}$, in which, by taking derivative, the DC component becomes zero and the oscillatory components remain. Thus the radial equation simplifies to 
\begin{equation}
-\frac{3a}{4r_{12}}\left[\frac{N^{+}_{12}A^{+}_{12}}{r_{12}^2}+R_2\sin(\phi_1-\phi_2)\right]\dot{\phi}_2=-\frac{\lambda}{\zeta}(R_1-R_0).
\label{ho:simplified}
\end{equation}
From section~\ref{SI_derivation}, recall the definition for $r_{12},A^{\pm}_{ij},N^{\pm}_{ij}$ in Eq.~\eqref{fulleq:def1}, \eqref{fulleq:def2} and \eqref{fulleq:def3}. We set $R_1=R_2=R_0$ in those expressions by taking the leading order DC component of radius (assume the radius $R_1$ is dominated by $R_0$ and next order is small), then they become
\begin{align}
\begin{split}
r_{12}^2 &= l^2-2lR_0(\cos\phi_1-\cos\phi_2)+2R_0^2(1-\cos\delta),
\\
A^{+}_{12}&=lR_0\sin\phi_2+R_0^2\sin\delta, \\
N^{+}_{12}&= -l\cos\phi_1+R_0(1-\cos\delta).
\end{split}
\end{align}
Expanding $1/r_{12}^3$ and $1/r_{12}$ using Taylor series yields
\begin{align}
\begin{split}
\frac{1}{r^3_{12}} &= \frac{1}{l^3}\left[1-\left(\frac{2R_0}{l}(\cos\phi_1-\cos\phi_2)-\frac{2R_0^2}{l^2}(1-\cos\delta)\right)\right]^{-\frac{3}{2}} \\
&= \frac{1}{l^3}\left(1+\frac{3R_0}{l}(\cos\phi_1-\cos\phi_2)-\frac{3R_0^2}{l^2}(1-\cos\delta)+\frac{15R_0^2}{2l^2}(\cos\phi_1-\cos\phi_2)^2+O\left(\frac{R_0^3}{l^3}\right)\right),
\\
\frac{1}{r_{12}} &= \frac{1}{l}\left[1-\left(\frac{2R_0}{l}(\cos\phi_1-\cos\phi_2)-\frac{2R_0^2}{l^2}(1-\cos\delta)\right)\right]^{-\frac{1}{2}}\\
&= \frac{1}{l}\left(1+\frac{R_0}{l}(\cos\phi_1-\cos\phi_2)-\frac{R_0^2}{l^2}(1-\cos\delta)+\frac{3R_0^2}{2l^2}(\cos\phi_1-\cos\phi_2)^2+O\left(\frac{R_0^3}{l^3}\right)\right).
\end{split} \label{ho:1onr}
\end{align}
The terms of order $O(R^3/l^3)$ in Eq.~\eqref{ho:1onr} are dropped. Next we make note of the following identity for later convenience:
\begin{align}
(\cos\phi_1-\cos\phi_2)^2
&= 1+\cos\sigma(\cos\delta-1)-\cos\delta.
\end{align}
We now consider the terms on the LHS of Eq.~\eqref{ho:simplified} one by one. First we look at the second term, utilising the Taylor expansion in Eq.~\eqref{ho:1onr} we have
\begin{align}
R_0\frac{\sin\delta}{r_{12}} &\approx \frac{R_0\sin\delta}{l}\left(1+\frac{R_0}{l}(\cos\phi_1-\cos\phi_2)-\frac{R_0^2}{l^2}(1-\cos\delta)+\frac{3R_0^2}{2l^2}(\cos\phi_1-\cos\phi_2)^2\right).
\end{align}
From here on we use the fact that when the intrinsic frequencies $\omega_1$ and $\omega_2$ are dominating the dynamics, they have similar magnitude but opposite sign (we use $\omega_1=-\omega_2=100\pi\text{ rad}\cdot$s$^{-1}$ in this paper), and the rate of change of $\delta\sim 2\omega_1t$ is much faster than $\sigma\sim 0$. Under the time scale that is large for changes in $\delta$ but small for changes in $\sigma$, the resulting average of $\sin\delta$, $\cos\delta\sin\delta$, $\sin\delta\cos\phi_{1,2}$ are all zero and we can treat any function of $\sigma$ as approximately constant. Therefore we have
\[
\left\langle R_0\frac{\sin\delta}{r_{12}}\right\rangle \approx 0.
\]
We need only focus on the other term on the LHS in Eq.~\eqref{ho:simplified}
\begin{align}
\begin{split}
\frac{N^{+}_{12}A^{+}_{12}}{r_{12}^3} &\approx
(-l\cos\phi_1+R_0(1-\cos\delta))(lR_0\sin\phi_2+R_0^2\sin\delta)\\
\quad
&\frac{1}{l^3}\left(1+\frac{3R_0}{l}(\cos\phi_1-\cos\phi_2)-\frac{3R_0^2}{l^2}(1-\cos\delta)+\frac{15R_0^2}{2l^2}(\cos\phi_1-\cos\phi_2)^2\right).
\end{split}
\end{align}
By similar argument above, we may drop the terms multiplied by $\sin\delta$ as they average to 0. So
\begin{align}
\begin{split}
\left\langle \frac{N^{+}_{12}A^{+}_{12}}{r_{12}^3}\right\rangle &\approx
\frac{R_0}{l}\sin\phi_2\left(-\cos\phi_1+\frac{R_0}{l}(1-\cos\delta)\right)
\\
& \left(1+\frac{3R_0}{l}(\cos\phi_1-\cos\phi_2)-\frac{3R_0^2}{l^2}(1-\cos\delta)+\frac{15R_0^2}{2l^2}(\cos\phi_1-\cos\phi_2)^2\right)
\end{split}
\end{align}
Multiplying term by term, notice the first term is $-\sin\phi_2\cos\phi_1 = (\sin\delta-\sin\sigma)/2$, as usual we can drop $\sin\delta$ on average. Therefore,
\begin{align}
\begin{split}
&\left\langle -\frac{R_0}{2l}\sin\sigma \left(1+\frac{3R_0}{l}(\cos\phi_1-\cos\phi_2)-\frac{3R_0^2}{l^2}(1-\cos\delta)+\frac{15R_0^2}{2l^2}(\cos\phi_1-\cos\phi_2)^2\right)\right\rangle \\
&\approx -\frac{R_0}{2l}\sin\sigma\left(1+\frac{9R_0^2}{2l^2}-\frac{15R_0^2}{2l^2}\cos\sigma\right).
\end{split}\label{ho:term1}
\end{align}
For the second multiplication, we note that $\sin\phi_2$ oscillates slower than $\cos\delta$ and $\cos^2\delta$ and we always drop $O(R^4/l^4)$. So on average we only need to focus on
\begin{align}
\begin{split}
\left\langle\frac{R_0^2}{l^2}\sin\phi_2(1-\cos\delta) \left(\frac{3R_0}{l}(\cos\phi_1-\cos\phi_2)\right)\right\rangle &=\left\langle\frac{3R_0^3}{l^3}(1-\cos\delta)(\sin\sigma-\sin(2\phi_2))/2\right\rangle
\\
&= \frac{9R_0^3}{4l^3}\sin\sigma.
\end{split}\label{ho:term2}
\end{align}
Combining the DC contributions from Eq.~\eqref{ho:term1} and \eqref{ho:term2}, we obtain the net contribution
\begin{equation}
-\frac{R_0}{2l}\sin\sigma\left(1-\frac{15R_0^2}{2l^2}\cos\sigma\right).
\end{equation}
Substituting this back to Eq.~\eqref{ho:simplified}, we have the result for the next-to-leading order in $R_0/l$ for average radial dynamics (take $\dot{\phi}_2\approx\omega_2$ at leading order)
\begin{equation}
R_1 = R_0-R_0\frac{\rho\zeta\omega_2}{\lambda}\left(1-\frac{15R_0^2}{2l^2}\cos\sigma\right)\sin\sigma.
\end{equation}
The higher order term has a direct impact on the leading order equation Eq.~\eqref{eq:sigma_smalhydrostiff}. By Taylor expanding the radius at denominator of $\dot{\phi}_{1,2}=R_0\omega_{1,2}/R_{1,2}+\cdots$ and taking the first order correction, we arrive at the modification for $\sigma$ shown in Eq.~\eqref{eq:sigma_bifurcation}.

\newpage
\section{Prediction of Rotor's Phase Speed} \label{SI_phase_speed}

Assume the system ($\theta = 0$ for simplicity) is in equilibrium and both oscillators are beating, hence $\sigma$ is constant and $\delta$ follows 
\begin{equation}
\dot{\delta} = A-B\sin\delta,
\label{pspeed:delta}
\end{equation}
where $|A|>|B|$. This can be solved by substitution and yields
\begin{equation}
\tan\left(\frac{\delta}{2}\right) = \frac{\Delta\tan u+B}{A}, \label{pspeed:tandelta}
\end{equation}
where $\Delta = \sqrt{A^2-B^2}$ and $u = \Delta t/2+c$ for some constant $c$. From this, we find that 
\begin{equation}
\sin\delta = \frac{\frac{\Delta}{A}\sin(2u)+2\frac{B}{A}\cos^2u}{1+\frac{B^2}{A^2}\cos(2u)+\frac{\Delta B}{A^2}\sin(2u)} .\label{pspeed:sindelta}
\end{equation}
Integrating Eq.~\eqref{pspeed:sindelta} over a period and taking the time average, we arrive at
\begin{equation}
\langle\sin\delta\rangle = \frac{A-\sqrt{A^2-B^2}}{B} = \frac{B}{2A}+O\left(\frac{B^3}{A^3}\right).
\label{pspeed:avg_sindelta}
\end{equation}
If we look back to Eq.~\eqref{eq:fibresonly} for $\sigma$ and $\delta$, we can substitute the corresponding $A$ and $B$ and obtain the leading order phase speed for $\phi_1$ (for $\phi_2$ it is opposite sign). In either the AP or IP state this is given by
\begin{align}
\begin{split}
\langle\dot{\phi}_1\rangle &= \langle\omega_1 - \frac{s^2}{2\zeta R_0^2}\left(k_+\sin\delta+k_-\cos\sigma\sin\delta\right) \rangle\\
& \approx \omega-\frac{\omega}{2}\left(\frac{s^2}{2\zeta\omega R_0^2}\right)^2(k_++k_-\cos\sigma)^2
\end{split}
\label{pspeed:fibres_only}
\end{align}
where $(k_+,k_-) = (k_y+k_x,k_y-k_x)$. Equation~\eqref{pspeed:fibres_only} also suggests the amplitude of instantaneous oscillation of phase speed is given by $ |\frac{s^2}{2\zeta R_0^2}\left(k_++k_-\cos\sigma\right)|$ and the oscillation always crosses the intrinsic speed $\omega$.

The implications of this formula are that in AP ($\sigma=0$) and IP ($\sigma = \pi$), the phase speed correction is proportional to $k_y^2$ and $k_x^2$ respectively. Physically in AP and IP states, the springs in the $x$ and $y$ directions have achieved their equilibrium lengths, and thus do not contribute to the dynamics. Provided  $k_x \text{ (in IP)}>k_y\text{ (in AP)}$, average phase speed is lower in IP but accompanies a larger oscillation. If $k_x$ is fixed and $k_y$ is changed to achieve transition, the phase speed in the IP state is lower than in the AP state. However, if $k_y$ is fixed and $k_x$ is changing, the situation is reversed. \\

Equations~\eqref{pspeed:avg_sindelta} and \eqref{pspeed:fibres_only} compare very well with numerical simulations. For $k_x=0.015 \ \text{Nm}^{-1}$, $k_y=0.006 \ \text{Nm}^{-1}$, and other parameters as in Table~\ref{table:params}, the estimate from simulations yields phase speed 42.071 Hz. Equation~\eqref{pspeed:avg_sindelta} predicts 42.071 Hz, and the leading order gives 42.070 Hz. For $k_x=0.008  \ \text{Nm}^{-1}$ and $k_y=0.015  \ \text{Nm}^{-1}$, simulations predict 34.672 Hz, Eq.~\eqref{pspeed:avg_sindelta} gives 34.672 Hz and the leading order gives 37.022 Hz. \\

We can apply similar reasoning to the case in Eq.~\eqref{eq:smallhydro}, in which the beads are also coupled through hydrodynamic interactions. The leading order for $\delta$, analogous to $\sigma$, is given by
\begin{equation}
\dot{\delta} = 2\omega+2\rho\omega\cos\sigma+O\left(\rho\omega\sin\delta\right)+\text{the internal coupling terms for } \delta \text{ in Eq.~\eqref{eq:fibresonly}}. \label{pspeed:delta_leading}
\end{equation}
Applying Eq.~\eqref{pspeed:avg_sindelta} using this modified definition for $A$ and $B$ gives the phase speed for $\phi_1$ in AP or IP with hydrodynamic interactions present in the limit $R_0\ll l$:
\begin{equation}
\langle\dot{\phi}_1\rangle  \approx \omega+\omega\rho\cos\sigma
-\frac{\omega}{2}\left(\frac{s^2}{2\zeta\omega R_0^2}\right)^2(k_++k_-\cos\sigma)^2. \label{pspeed:hydro}
\end{equation}
The expression consists of the basal coupling term in Eq.~\eqref{pspeed:fibres_only}, together with a hydrodynamic correction. This hydrodynamic correction enhances the phase speed for the AP state and reduces it in the IP state, as expected physically \cite{Leptos:2013}.

For the physical parameters used in the paper (Table~\ref{table:params}), the basal coupling term is dominating Eq.~\eqref{pspeed:hydro} as $\rho\approx 0.02$ is small. If the system is achieving a transition in synchronization states (IP $\leftrightarrow$ AP) through changes in just one of $(k_x,k_y)$, to be consistent with the experimental observation of significant changes in phase speed, the model suggests $k_y$ is the primary variable basal coupling.

\begin{figure}[htp]
\begin{center}
\includegraphics[width=0.55\columnwidth]{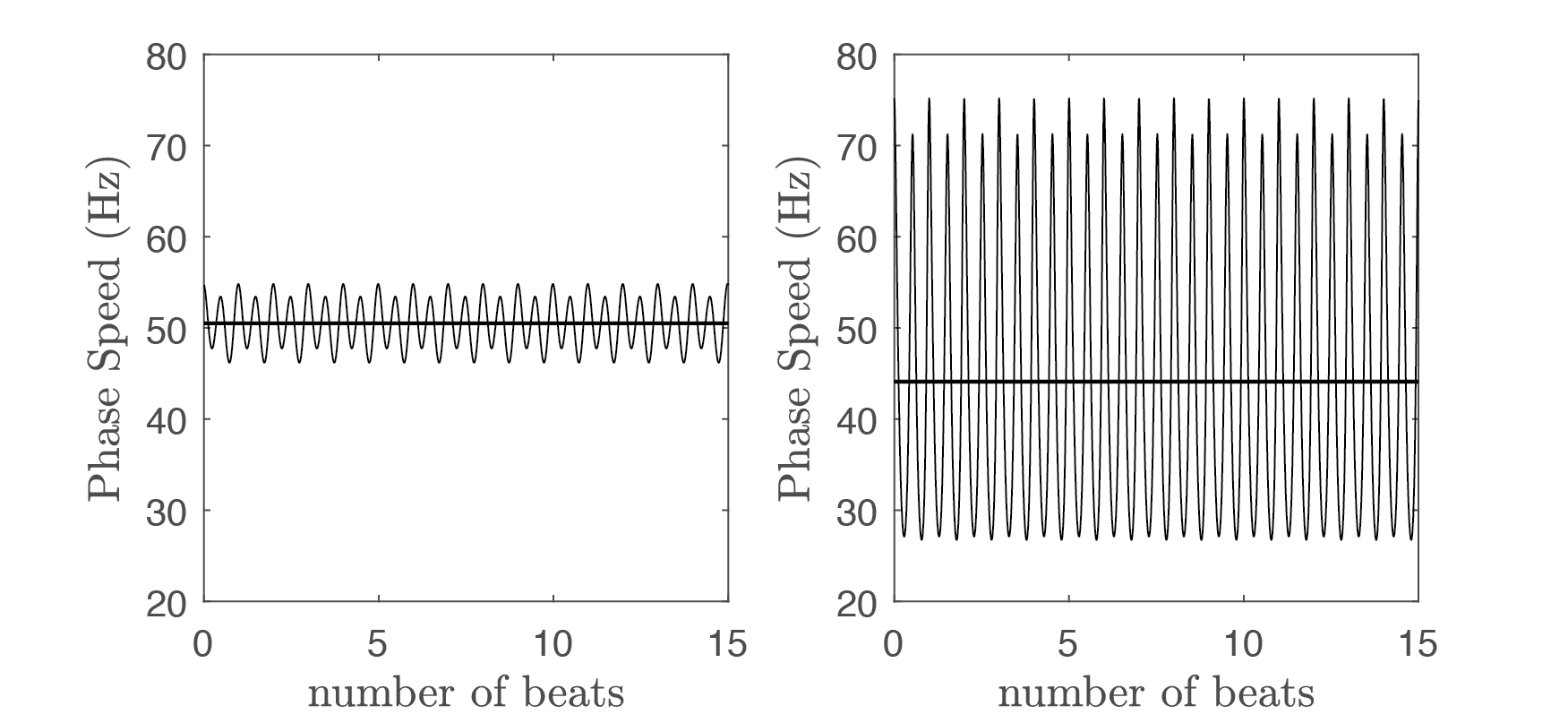}
\end{center}
\caption{Instantaneous phase speed profile with hydrodynamics present. Black bold lines indicate the average phase speed. (left) $k_x = 0.005$ Nm$^{-1}$ and $k_y=0$ Nm$^{-1}$ in AP. (right) $k_x = 0.005$ Nm$^{-1}$ and $k_y=0.02$ Nm$^{-1}$ in IP.}
\label{FIG_S1}
\end{figure}

In the simulations presented in the paper (with hydrodynamics), where we only vary $k_y$, we observed the phase speed is lower in the IP state compared to the AP state. For parameters used in Table~\ref{table:params} and $k_x = 0.005$ Nm$^{-1}$, the phase speed reduction between $k_y=0$ Nm$^{-1}$ and $k_y=0.02$ Nm$^{-1}$ is about 12.7\% (see Fig.~\ref{FIG_S1}). The formula Eq.~\eqref{pspeed:hydro} predicts a reduction of about 13.6\%.

\section{Existence of larger excursion modes} \label{SI_larger_excursions}

\begin{figure}[htp]
\begin{center}
\includegraphics[width=0.5\columnwidth]{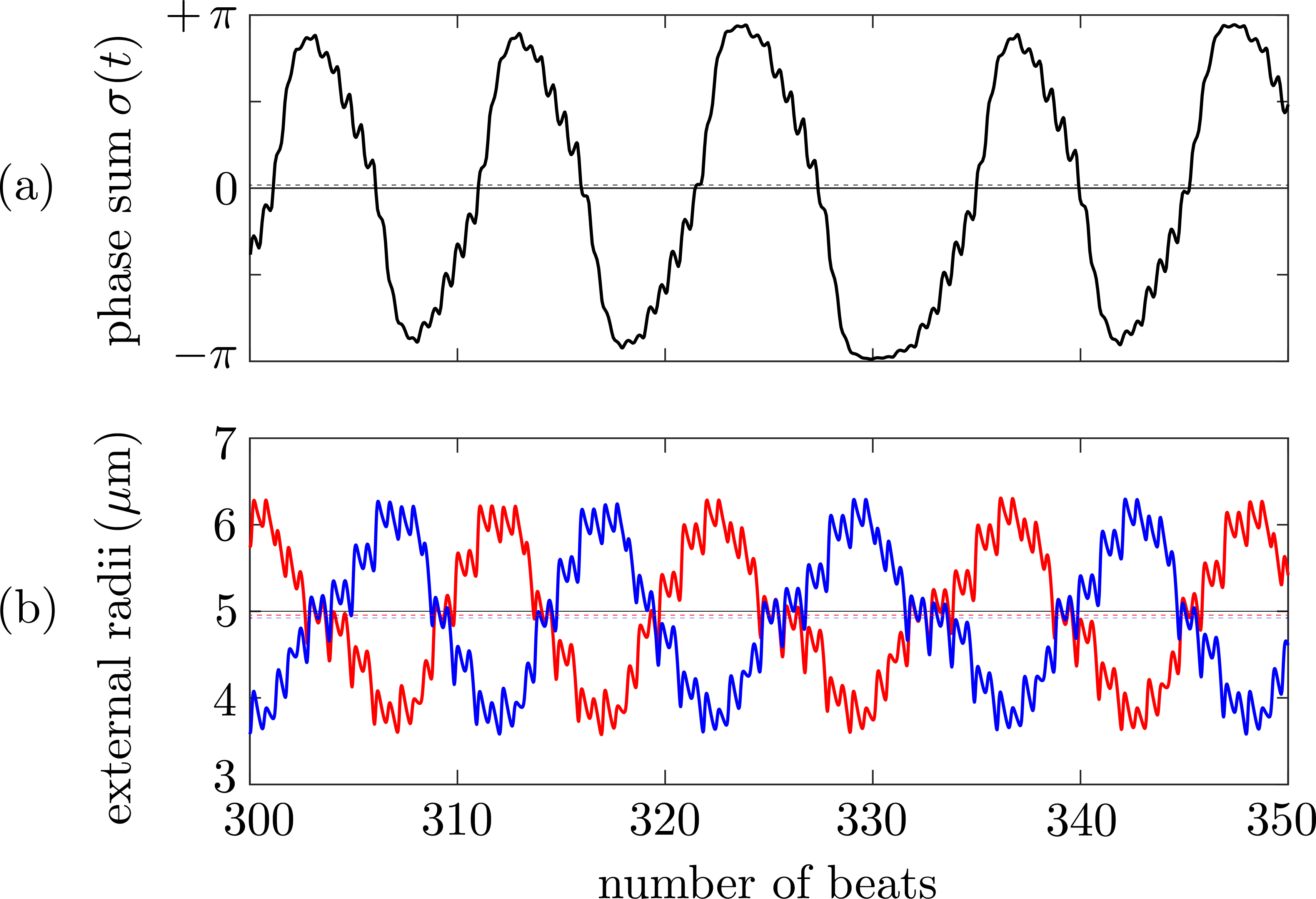}
\end{center}
\caption{Deterministic switching between IP and AP states. (a) Phase sum $\sigma(t) = \phi_1(t) + \phi_2(t)$ and (b) external rotor radii, $R_i(t)$. Parameters are as in Table~\ref{table:params}, except with $a=1 \ \mu$m, $k_x = 5 \times 10^{-3} \ \text{Nm}^{-1}$, and $k_y = 7.65 \times 10^{-3} \ \text{Nm}^{-1}$.}
\label{FIG_S2}
\end{figure}

In section~\ref{Section_full_model} of the main paper, Fig.~\ref{FIG5}a,b demonstrates that for intermediate values of $k_y$, the phase sum $\sigma(t)$ oscillates with large amplitude but zero mean, on a timescale that is long compared to the individual rotor period (see panel (iii)). Figure~\ref{FIG_S2} illustrates the effect of increasing the radius, $a$, of the external bead. The chosen value $k_y = 7.65 \times 10^{-3} \ \text{Nm}^{-1}$ is just below the symmetry breaking threshold. For these parameters, the phase sum oscillates between $\approx \pi$ and $-\pi$ with a period of $\approx 12$ beats. While $\langle \sigma(t) \rangle = 0$ would suggest an AP state, this system actually spends considerable time close to IP states ($\sigma = \pm \pi$), with rapid swings past the $\sigma =0$ states. 

\newpage
\section{Coexistence of synchronized states} \label{SI_coexistence}

Figure~\ref{FIG5}c,e in the main text shows that for $l \leq 35 \ \mu$m, the full system displays a discontinuous transition in $\langle \sigma(t) \rangle$ as $k_y$ is increased. For each of the numerical simulations used to produce Fig.~\ref{FIG5}c,e, the same initial conditions were used, $(\phi_1, \phi_2) = (0, \pi/2)$. In this section, we further examine the nature of this discontinuity, and explore the capacity for the system to support alternative values of $\langle \sigma(t) \rangle$ for the same $(k_x,k_y)$.

\begin{figure}[htp]
\begin{center}
\includegraphics[width=0.30\columnwidth]{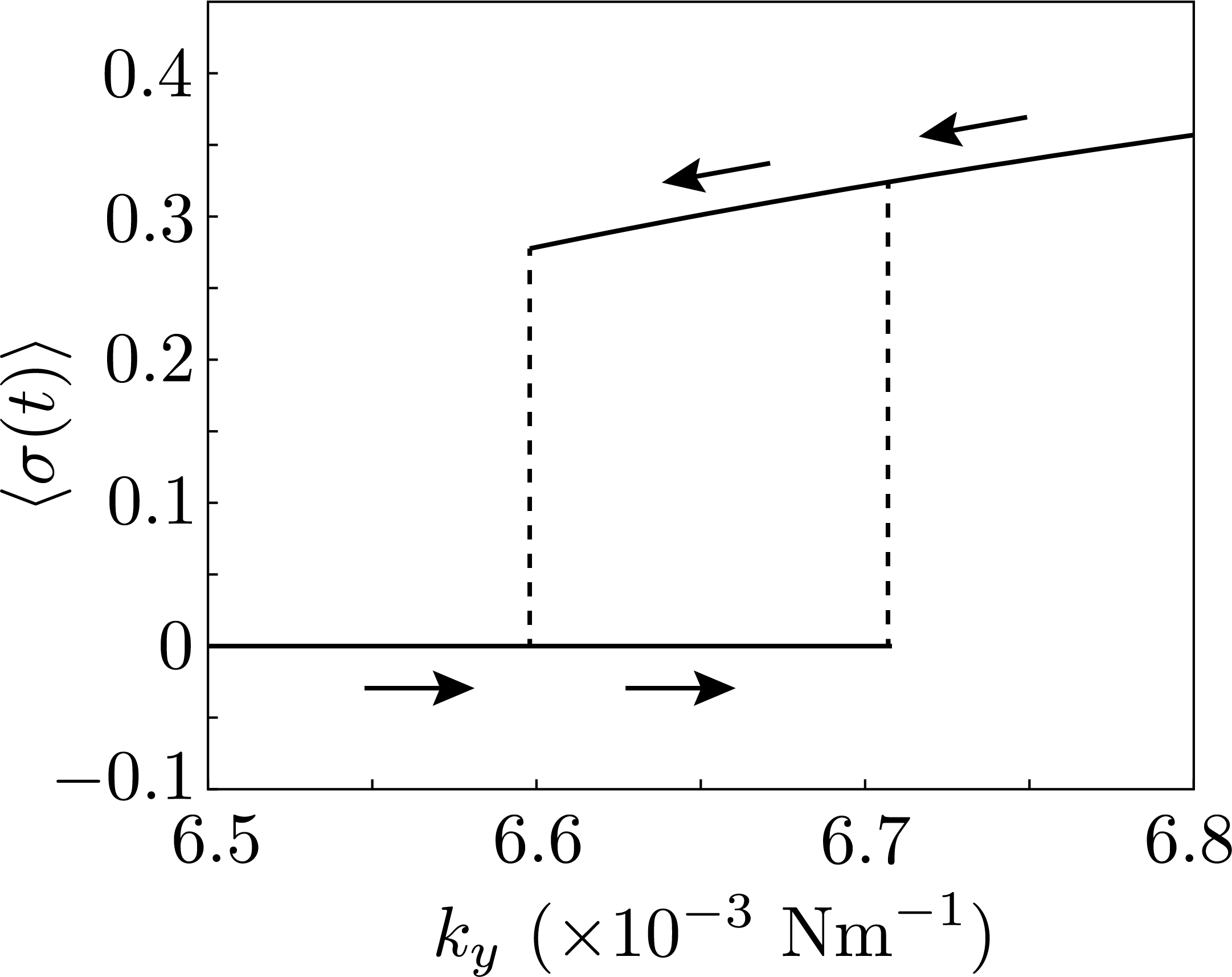}
\end{center}
\caption{Bifurcation diagram showing $\langle \sigma(t) \rangle$ as a function of $k_y$, for $l=15 \ \mu$m and $k_x = 5 \times 10^{-3} \ \text{Nm}^{-1}$. Symmetric negative $\langle \sigma(t) \rangle$ branch not shown. All other parameters are as in Table~\ref{table:params}.}
\label{FIG_S3}
\end{figure}

For the {\it Chlamydomonas}-like separation of $l=15\,\mu$m, we start with a value of $k_y = 6.5 \times 10^{-3} \ \text{Nm}^{-1}$, for which $\langle \sigma(t) \rangle$ converges to zero (AP). Once the steady state is achieved, we gradually increase the value of $k_y$, on a timescale much longer than all other timescales in the system. The system continues to support AP synchronization until $k_y \simeq 6.7 \times 10^{-3} \ \text{Nm}^{-1}$, at which point a discontinuous jump brings the system to another stable point with $\langle \sigma(t) \rangle =\pm 0.32$ (see Fig.~\ref{FIG_S3}). Similarly, starting instead at $k_y = 6.8 \times 10^{-3} \ \text{Nm}^{-1}$ along the positive $\langle \sigma(t) \rangle$ branch, we can gradually decrease the $k_y$ value keeping on the same branch until $k_y \simeq 6.6 \times 10^{-3} \ \text{Nm}^{-1}$, at which point  the average phase sum jumps to $\langle \sigma(t) \rangle=0$. As Fig.~\ref{FIG_S3} illustrates, there is clearly a small but finite interval in $k_y$ where the system displays 3-state multi-stable dynamics (symmetric negative branch of 
$\langle \sigma(t) \rangle$ not shown). In this small region, the relative sizes of the basins of attraction for the different states depend on the value of $k_y$.

The region of parameter space in which this coexistence occurs is relatively small compared to the overall transition zone presented in Fig.~\ref{FIG5}c ($6.6 \times 10^{-3} \ \text{Nm}^{-1} < k_y < 17 \times 10^{-3} \ \text{Nm}^{-1}$). Moreover, the width of the coexistence zone diminishes with increasing separation $l$ and eventually disappears for $35\,\mu\text{m}<l<50\,\mu\text{m}$, when the bifurcation changes nature. As the 3-states coexistence region is narrow, it seems more likely that changes in the synchronization state would be mediated through underlying changes in the basal body stiffness, as outlined in the main text, rather than stochastic jumping between the coexisting states for a fixed $k_y$.

\end{document}